\documentclass{aa}
\usepackage{txfonts}
\usepackage{graphicx}
\begin{document}

\newcommand{\beqa}{\begin{eqnarray*}}
\newcommand{\eeqa}{\end{eqnarray*}}
\newcommand{\beqan}{\begin{eqnarray}}
\newcommand{\eeqan}[1]{\label{#1}\end{eqnarray}}
\newcommand{\beq}{\begin{equation}}
\newcommand{\eeq}{\end{equation}}
\newcommand{\diff}{{\rm d}}
\newcommand{\drr}{\frac{\partial}{\partial r}}
\newcommand{\dtt}{\frac{\diff}{\diff t}}
\newcommand{\dr}[1]{\frac{\partial  #1}{\partial r}}
\newcommand{\dt}[1]{\frac{\partial  #1}{\partial t}}
\newcommand{\lp}{ \left(}
\newcommand{\rp}{ \right)}
\newcommand{\lc}{ \left[}
\newcommand{\rc}{ \right]}
\newcommand{\cf}{{\it cf.}~}
\newcommand{\ie}{{\it i.e.}~}
\newcommand{\eg}{{\it e.g.}~}

\def\lta{\mathrel{{\lower 3pt\hbox{$\mathchar"218$}} \hspace{-8pt}
     \raise 2.0pt\hbox{$\mathchar"13C$} \hspace{-5pt} }}
\def\gta{\mathrel{{\lower 3pt\hbox{$\mathchar"218$}} \hspace{-8pt}
     \raise 2.0pt\hbox{$\mathchar"13E$} \hspace{-5pt} }}

\bibliographystyle{plain}

\title{Hydrodynamical stellar models including rotation, \\ internal gravity waves 
and atomic diffusion}
\subtitle{I. Formalism and tests on Pop I dwarfs}

\author{Suzanne Talon\inst{1} and Corinne Charbonnel\inst{2,3}}

\offprints{Suzanne Talon}

\institute{
D\'epartement de Physique, Universit\'e de Montr\'eal, Montr\'eal PQ H3C 3J7, Canada
\and Observatoire de Gen\`eve, 51, ch. des Maillettes, 1290 Sauverny, Switzerland 
\and Laboratoire d'Astrophysique de Toulouse et Tarbes, CNRS UMR 5572, OMP,
14, Av. E.Belin, 31400 Toulouse, France \\ 
(Suzanne.Talon@astro.umontreal.ca, Corinne.Charbonnel@obs.unige.ch)}

\date{Accepted for publication}

\authorrunning{S. Talon \& C. Charbonnel}
\titlerunning{Stellar models including rotation and IGW I.}

\abstract{In this paper, we develop a formalism in order to incorporate the contribution
of internal gravity waves to the transport of angular momentum and chemicals over
long time-scales in stars.
We show that the development of a double peaked shear layer acts as a filter for waves, and how the asymmetry 
of this filter produces momentum extraction from the core when it is rotating faster than the
surface. Using only this filtered flux, it is possible to follow the contribution
of internal waves over long (evolutionary) time-scales. 

We then present the evolution of the internal
rotation profile using this formalism
for stars which are spun down via magnetic torquing.  We 
show that waves tend to slow down the core, creating a
``slow" front that may then propagate from the core to the surface. 
Further spin down of the surface leads to
the formation of a new front. 
Finally we show how this momentum transport reduces rotational mixing in a 
$1.2\,M_\odot$, $Z=0.02$ model, leading to a surface lithium abundance 
in agreement with observations in the Hyades.
}

\maketitle

\section{Introduction}

Stellar models are getting more and more sophisticated.
In order to explain detailed observed features of stars 
in various places of the Hertzsprung-Russell diagram, 
modern stellar evolution codes must indeed incorporate several 
complex physical processes which are often referred to as
``non-standard''. 
The main ones are :
\begin{itemize}
\item 
Atomic diffusion (gravitational settling, thermal diffusion,
radiative forces);
\item Large scale mixing due to rotation (turbulence, meridional circulation);
\item Convective overshooting;
\item Internal gravity waves (IGW);
\item Magnetic fields.
\end{itemize}

All these processes are not necessarily present at the same time 
everywhere in the HR diagram. 
Additionally they act on very different characteristic timescales 
in stars of various initial masses and evolutionary stages. 
In order to provide a coherent picture of stellar evolution,
one must thus understand why some dominate in certain stellar types
and not in others. 
Furthermore, one has to pay attention to their 
possible interactions and interdependence. 

These mechanisms affect evolutionary tracks,
lifetimes, surface abundances, chemical yields etc.
Their impact on the stellar structure and evolution arises mainly 
from the redistribution of chemical elements that they cause, and 
which is related (except in the case of atomic diffusion) to the 
induced redistribution of angular momentum inside the stars. 
During the last fifteen years, many studies have been devoted 
to describe the evolution of the angular momentum distribution
because this pattern governs the extent of rotation-induced mixing in stellar 
interiors.
For models in which the
internal rotation law evolves under the effects of meridional circulation, shear 
mixing, horizontal turbulence, mass loss, contraction and expansion (i.e., neglecting
the transport by IGW and by magnetic fields) 
the main results can be summarized as follows :
\begin{itemize}
\item
On the one hand, models which take into account the 
hydrodynamical processes induced by rotation as described by
Zahn (1992), Maeder (1995), Talon \& Zahn (1997) and Maeder \& Zahn (1998)
are very successful in explaining the observed ``anomalies'' in 
stars which do not have extended envelopes. Among the many 
successes of these models one can quote the reproduction 
of the left side of the Li dip in field and open cluster stars, of the 
constancy of the CNO abundances within the Li dip  
and of the evolution of Li abundance in sub-giants (Charbonnel \& Talon 1999, 
Palacios et al. 2003, Pasquini et al.2004). 
For more massive stars these models explain for example the observed 
He and N enrichment in main sequence O- and early B-type stars, 
in OB super-giants and in A-type SMC super-giants. They also account for the observed 
variations of the Wolf-Rayet star populations as well as the fractions 
of type Ib/Ic supernovae with respect to type II SN at various metallicities 
(see Maeder \& Meynet 2000, Meynet \& Maeder 2005 and references therein).
\item
On the other hand, the same input physics fails to reproduce some of the most
constraining observed features in low-mass stars which have 
deep convection envelope. First of all, meridional circulation and 
shear turbulence alone are not efficient enough to enforce the flat solar
rotation profile measured by helioseismology (Brown et al. 1989;
Kosovichev et al. 1997; Couvidat et al. 2003; Pinsonneault et al. 1989;
Chaboyer et al. 1995 
\footnote{Note that the Yale group computed the evolution of angular momentum
using a simplified description of the action of meridional circulation which
was considered as a diffusive rather than as an advective process}
; Matias \& Zahn 1998). Additionally the rise 
of the Li abundance on the red side of the dip cannot be explained 
if one assumes that momentum is transported only by the 
wind-driven meridional circulation in those main-sequence stars which 
are efficiently spun down via magnetic torquing
(Talon \& Charbonnel 1998).
Last but not least,
the current rotating models are insufficient to explain the anomalies in 
C and N isotopes observed in field and cluster red giant stars 
(Palacios et al. 2005).
\end{itemize}

The successes and difficulties described above have revealed the occurrence of 
an additional process that participates in the transport of angular momentum
in relatively low-mass stars which have extended convective envelopes and 
are spun down via magnetic braking in their early evolution 
(Talon \& Charbonnel 1998).
This process must act in conjunction with meridional circulation, turbulence and
atomic diffusion. As of now, two candidates have received some attention, namely
internal gravity waves and magnetic fields.

Internal gravity waves have initially been invoked as a source of mixing
for chemicals (Press 1981, Garc\'{\i}a L\'opez \& Spruit 1991,
Schatzman 1993, Montalb\'an 1994,
Montalb\'an \& Schatzman 1996, 2000, Young et al. 2003).
Ando (1986) studied the transport of momentum
associated with standing, gravity waves. He showed how momentum redistribution 
by these waves may increase
the surface velocity to induce episodic mass-loss in Be stars.
Goldreich \& Nicholson (1989) used them later in order to explain
the evolution of the velocity of binary stars, producing  
synchronization that proceeds
from the surface to the core. Traveling internal gravity 
waves have since been invoked as an
important process in the redistribution of angular momentum
in single stars spun down by magnetic torquing (Schatzman 1993, 
Kumar \& Quataert 1997, Zahn, Talon, \& Matias 1997).

In a previous series of papers, we examined the generation of 
internal gravity waves by the surface convection zone of stars
with various masses and metallicities. We found that these waves, which 
are able to extract angular momentum from the deep solar interior 
(Talon, Kumar, \& Zahn 2002, hereafter TKZ), have a 
very peculiar mass (or more precisely effective temperature) dependence
and could possibly dominate the transport of angular momentum 
in stars with deep enough convective envelope (Talon \& Charbonnel 2003,
2004). We suggested that such a dependence could lead to a
coherent picture of rotational mixing in stars of all masses at various
evolution phases. It could for example explain simultaneously the cold side 
of the Li dip as well as the solar rotation profile and the existence 
of fast rotating horizontal branch stars 
(see Talon \& Charbonnel 2004 for more details). 

An important characteristic of internal waves is that, unless
they are damped, they conserve their angular momentum even when
their local frequency is modified by Doppler shifting. This should be
kept in mind throughout the reading of this paper.
For a comprehensive review of gravity wave properties, we suggest 
Bretherton (1969) and Zahn et al. (1997) for an application to the stellar,
spherical case.

The other mechanism that has been invoked to enforce the Sun's flat
rotation profile requires a pre-existing fossil magnetic field
(Charbonneau \& Mac Gregor 1993; Barnes, Charbonneau, \& Mac Gregor 1999). 
However no mass dependence is expected in that case. 
This is in contradiction
with the Li dip constraint (although in this case this feature could be 
explained in purely solid body rotation by the combined action 
of solid body meridional circulation and radiative forces; see Charbonneau 
\& Michaud 1988) and more importantly with the large rotational velocities 
measured in some horizontal branch stars. 

\begin{figure*}[t]
\centering {
\includegraphics[width=13cm]{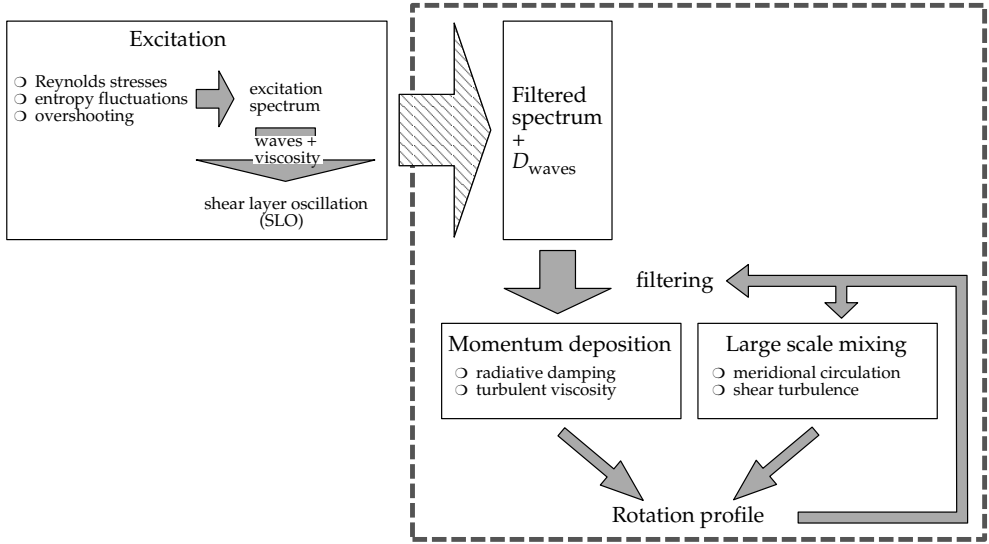}
}
\caption{Schematic view of the physical processes and their interactions involved
in momentum redistribution by IGW. Complete calculations take into account all 
processes enclosed in the dashed box.
\label{schema}}
\end{figure*}

Let us also mention calculations of massive star models
by Maeder \& Meynet (2003) in which 
the Taylor-Spruit dynamo (Spruit 1999, 2002), which is thought to be the most efficient
dynamo mechanism in the radiative region, is included. When considered as is,
this mechanism induces almost solid body rotation, efficiently reducing the extent 
of rotational mixing. It also leads to stellar models whose properties are 
closer to those of standard models and thus, in lesser agreement with
stellar observations. When reviewing the efficiency of this mechanism by taking
into account energy conservation, it is found that the dynamo is not
as efficient as first thought. 
In fact, 
calculations for a $15~M_{\odot}$ star model show that adding the contribution 
of magnetic field only 
slightly modifies the results obtained with the 
purely hydrodynamic
models (Maeder \& Meynet 2004 and Maeder, private communication).
Let us recall that for such massive stars with no convective envelope
internal gravity waves can not be produced efficiently and thus the present 
rotating models would remain unchanged in our global picture.\footnote{The potential effect 
of internal gravity waves produced by the convective stellar core should be
modest, although it remains to be quantified.}

All these results and constraints obtained for stars covering
a large fraction of the HR diagram
strongly incite further studies of the combined effects of rotation
and internal gravity waves in stars where these waves are known to 
be efficiently produced. Although they do not definitively disgrace 
magnetic fields which certainly play a role in the complete picture, 
they bring sufficient arguments to justify complete tests of a 
fully hydrodynamical model before including MHD effects. 

This is what we intend to do in the present series of papers where 
we will investigate the combined and sometimes conflicting effects 
of rotation, internal waves and atomic diffusion in low- 
and intermediate-mass stars at various stages of their evolution. 
We will present the first fully hydrodynamical stellar models that include 
self-consistently the chemical and momentum redistribution 
by these three mechanisms. 

In the present paper we discuss the formalism that we will use in the 
hydrodynamical models which will be presented here and in forthcoming studies. 
A schematic view of all physical ingredients 
included and their interactions is
shown in Fig.~\ref{schema}.
In \S~2,
we first describe the prescription we follow for wave generation or excitation. 
Then in \S~3, we come back to the properties of the shear layer oscillation (or SLO)
which builds up below the surface convection zone of the star 
and on the momentum extraction by the waves in the radiative region. 
We explain how an excited wave spectrum combined with the action of the SLO
may be replaced by a filtered spectrum and a diffusion coefficient, as illustrated in
Fig.~\ref{schema}.
General equations for the transport of angular momentum and 
chemicals are written in the global scheme (\S~4 and 5), which takes into account
momentum deposition beyond the shear layer and momentum transport by meridional circulation and
shear turbulence. 
Then we recall some properties of the transport by waves using a static model
(\S~6). Finally we present the first results for a $1.2\,M_{\odot}$, 
$Z=0.02$ star computed within the complete physical framework (\S~7) 
before concluding (\S~8). 

\section{Internal Gravity Wave Generation \label{sec:waveexcitation}}

Gravity waves are produced, among other things, by the injection of kinetic
energy from a turbulent region to an adjacent stable region.
This is observed for example at the border of clouds in the
earth's atmosphere (Townsend 1965) and also in laboratory experiments
(Townsend 1958). It should also occur in stars.
This was already
illustrated in the early (2D) numerical simulations of convection including
penetration by Hurlburt et al. (1986, see also Hurlburt et al. 1994, 
Andersen 1994, Nordlund et al. 1996, Kiraga et al. 2000, Rogers \& 
Glatzmeier 2005). 
There are two ways to excite those waves:
\begin{itemize}
\item convective overshooting in a stable region;
\item bulk excitation, similar to that of the solar pressure waves.
\end{itemize}

Excitation by overshooting is probably the most difficult to evaluate analytically. The first 
attempt was made by Garc\'\i a L\'opez \& Spruit (1991) by assuming that the
pressure perturbation produced by turbulent eddies
at the radiative/convective boundary is equal to the wave pressure perturbation.
In this model, stochastic eddies of a given size are
considered to contribute to the excitation of a whole wavelength spectrum. As formulated, this 
model rests on the assumption of homogeneous turbulence, although it could be
modified to incorporate the presence of downdrafts observed in numerical
simulations of convection. Another estimate for this process has been made by
Fritts, Vadas, \& Andreassen (1998), in a study aimed at estimating the residual
circulation induced by latitude dependent dissipation in the tachocline. 
However, they concentrate on the small wavelength waves that dissipate 
close to the convection zone and their mechanism does not take into account the combination
of small scale eddies in order to produce low degree waves. Such waves are essential in
order to influence the inner layer on an evolutionary time-scale (TKZ).
Let us also mention here that the results presented in
this study depend on the latitudinal differential rotation and cannot (as of
now) be generalized to stars other than the Sun.

Kiraga et al. (2003) tried to evaluate the validity of the Garc\'\i a L\'opez \& Spruit
approach by
comparing the predictions of this model to a simulation of penetrative convection.
The peak wave spectrum produced by the simulation was similar in
amplitude to that of this parametric model. However, in the simulation
modes where excited over a
much broader range of frequencies and wavelengths. It is not clear whether this
is related to the bi-dimensional nature of the simulation of to shortcomings of
the model.

Internal gravity waves can also be excited in the convection zone itself. In that region, modes
are evanescent and their amplitude is proportional to $\exp \lc -\int \diff r \, k_r \rc$,
where $k_r$ is the radial wave number.
Press (1981) used the formalism of Goldreich \& Keeley (1977) to show that 
Reynolds-stress and buoyancy 
may excite gravity waves with a rather large amplitude at the bottom of the convection
zone.
Goldreich \& Kumar (1990) and Goldreich, Murray, \& Kumar (1994, GMK) completed the 
Goldreich \& Keeley formalism in order to apply it to solar p-modes. Their model quite
successfully reproduces the solar spectral energy input rate distribution, 
provided one free parameter which describes the geometry of turbulent eddies is
calibrated. In that case, driving is dominated by entropy fluctuations.
Balmforth (1992) made a similar study, using a somewhat different formalism. 
Subject to the calibration of a free parameter, he is also able to reproduce the spectral
energy distribution; however, it is the Reynolds-stress that is the main source of
driving.

\begin{figure}[t]
\centering {
\includegraphics[width=6.5cm,angle=270]{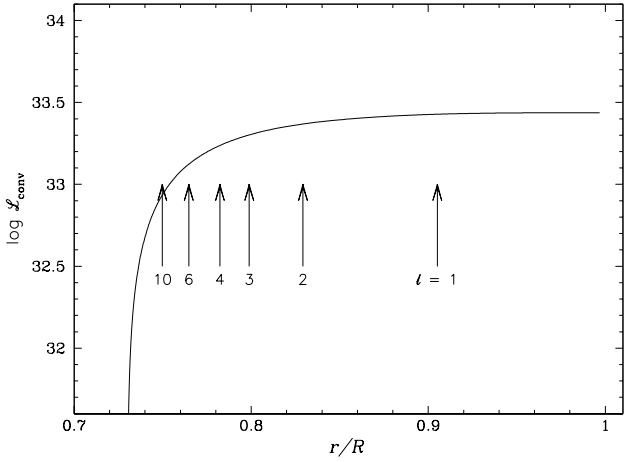}
}
\caption{Penetration of IGW into the convection zone of a ZAMS $1\,M_\odot$
model for various degrees $\ell$. The 
arrows indicate the 
depth where the amplitude is reduced by a factor of 2. 
Also shown is the local convective flux given by the mixing length theory.
\label{excitation}}
\end{figure}

In the present study, we follow Kumar \& Quataert (1997) and apply the GMK formalism to
traveling internal gravity waves. This was also used by Kumar, Talon, \& Zahn (1999),
TKZ, Talon \& Charbonnel (2003, 2004). 
The energy flux per
unit frequency ${\cal F}_E$ is then
\begin{eqnarray}
{\cal F}_E \lp \ell, \omega \rp &=& \frac{\omega^2}{4\pi} \int dr\; \frac{\rho^2}{r^2}
   \left[\left(\frac{\partial \xi_r}{\partial r}\right)^2 +
   \ell(\ell+1)\left(\frac{\partial \xi_h}{\partial r}\right)^2 \right]  \nonumber \\
 && \times  \exp\left[ -h_\omega^2 \ell(\ell+1)/2r^2\right] \frac{v^3 L^4 }{1
  + (\omega \tau_L)^{15/2}},
\label{gold}
\end{eqnarray}
where
$\xi_r$ and $[\ell(\ell+1)]^{1/2}\xi_h$ are the radial and horizontal
displacement wave-functions which are normalized to unit energy flux just
below the convection zone, $v$ is the convective velocity, $L$ is the radial
size of an energy bearing turbulent eddy, $\tau_L \approx L/v$ is the
characteristic convective time, and $h_\omega$ is the
radial size of the largest eddy at depth $r$ with characteristic frequency of
$\omega$ or greater ($h_\omega = L \min\{1, (2\omega\tau_L)^{-3/2}\}$).
The radial wave number is related to the horizontal wave number $k_h$ by
\beq
k_r^2 = \lp \frac{N^2}{\sigma^2} -1 \rp k_h^2 = 
\lp \frac{N^2}{\sigma^2} -1 \rp \frac{\ell \lp \ell +1 \rp}{r^2}
\eeq
where $N^2$ is the Brunt-V\"ais\"al\"a frequency. In the convection zone, the mode is
evanescent and the penetration depth varies as $\sqrt{\ell \lp \ell +1 \rp}$\footnote{This is
the theoretical dependence. Damping can be enhanced by turbulence, which would reduce
the surface amplitude of the mode (see \eg Andersen, 1996).}.
Figure~\ref{excitation} compares this distance for various modes with the 
convective energy that is
locally available. In the outer part of the model, the luminosity is carried almost 
uniquely by convection.

The momentum flux per unit frequency ${\cal F}_J$ is then related to the energy flux by                               
\begin{equation}                                                                          
{{\cal F}_J}\lp m, \ell, \omega \rp = \frac{m}{\omega} {\cal F}_E\lp \ell, \omega \rp  
\end{equation}   
(Goldreich \& Nicholson 1989, Zahn et al. 1997).

\section{Evolution of Angular Momentum}

\begin{figure*}[t]
\centering{
\includegraphics[width=5.5cm]{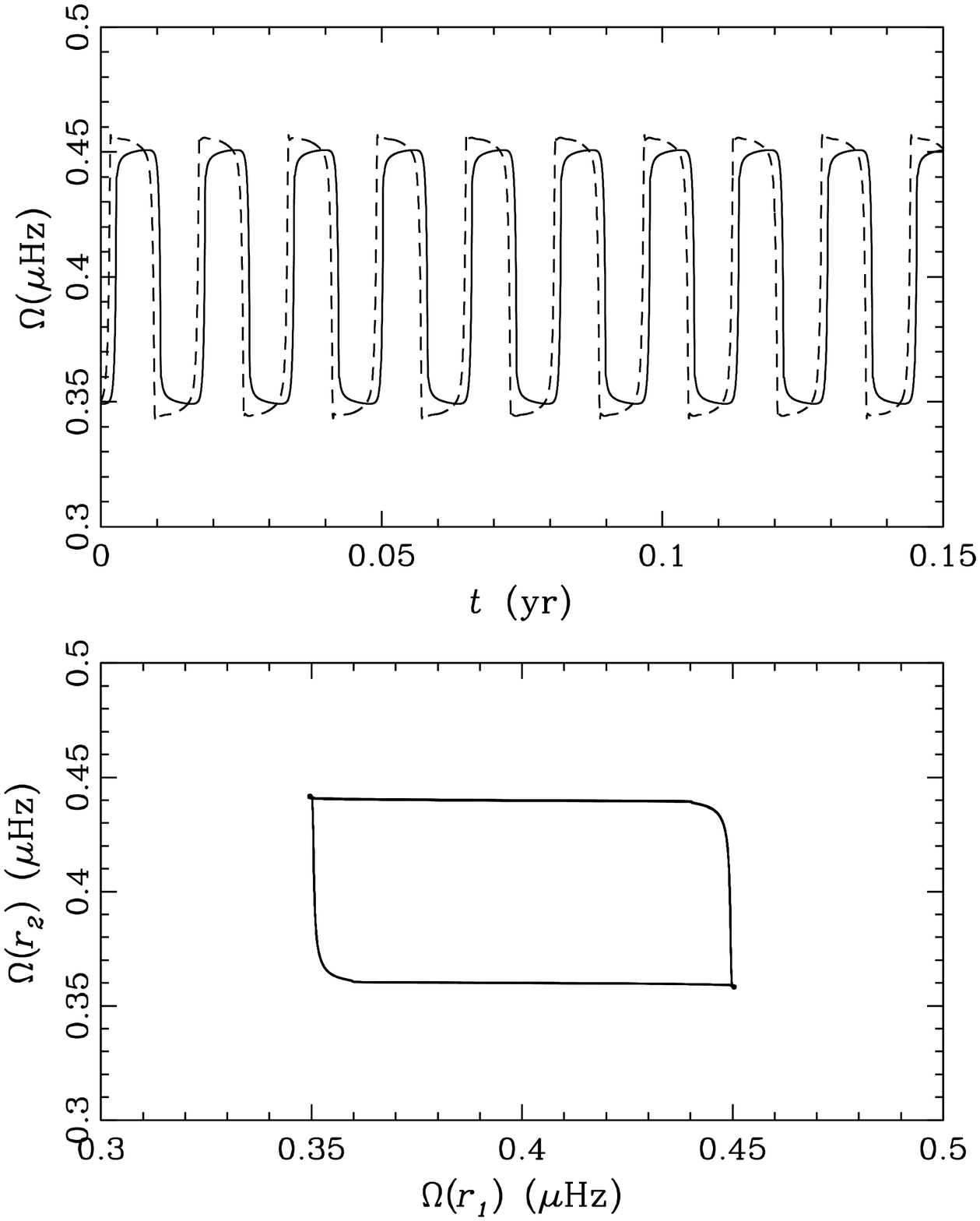} 
\includegraphics[width=5.5cm]{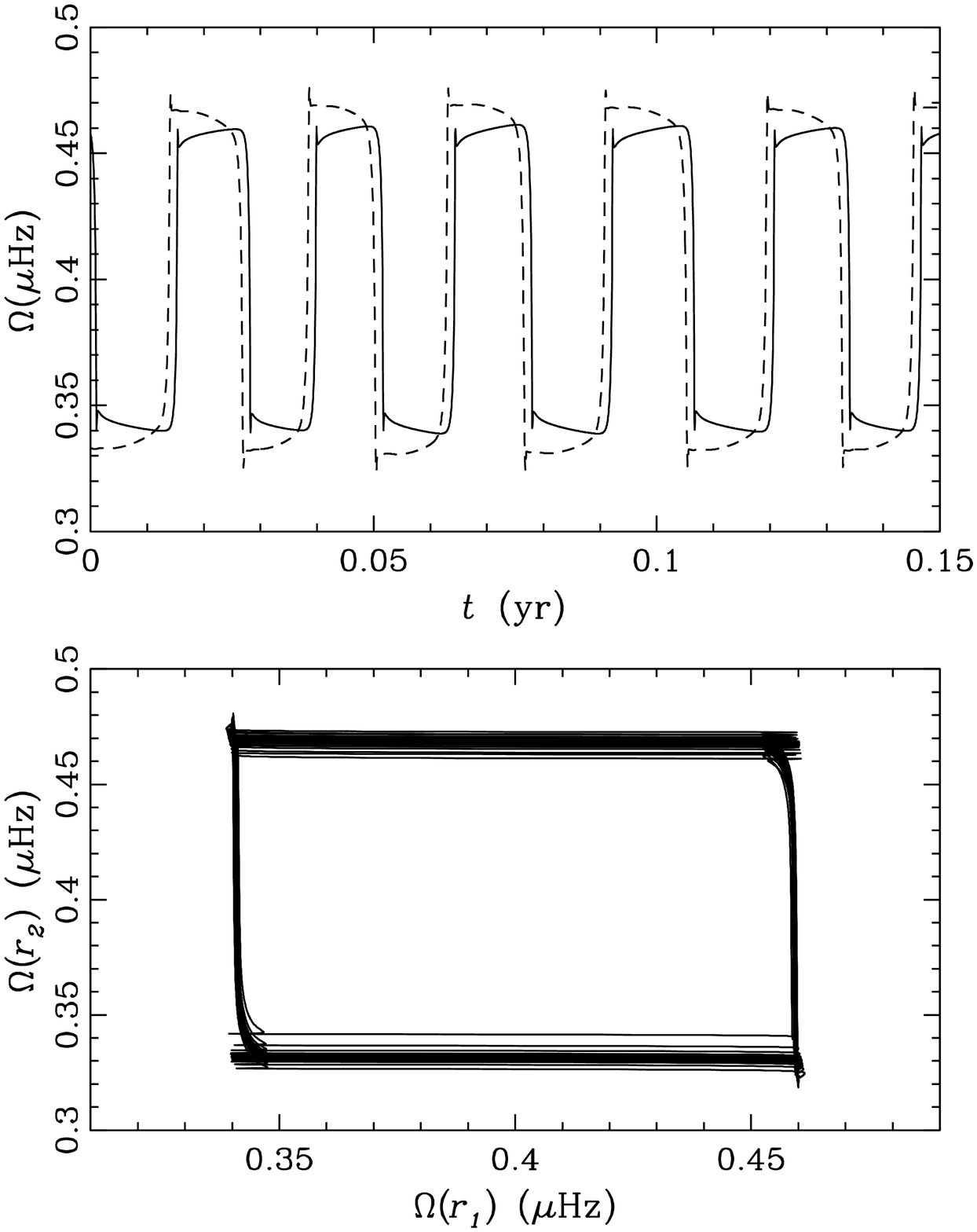}
\includegraphics[width=5.5cm]{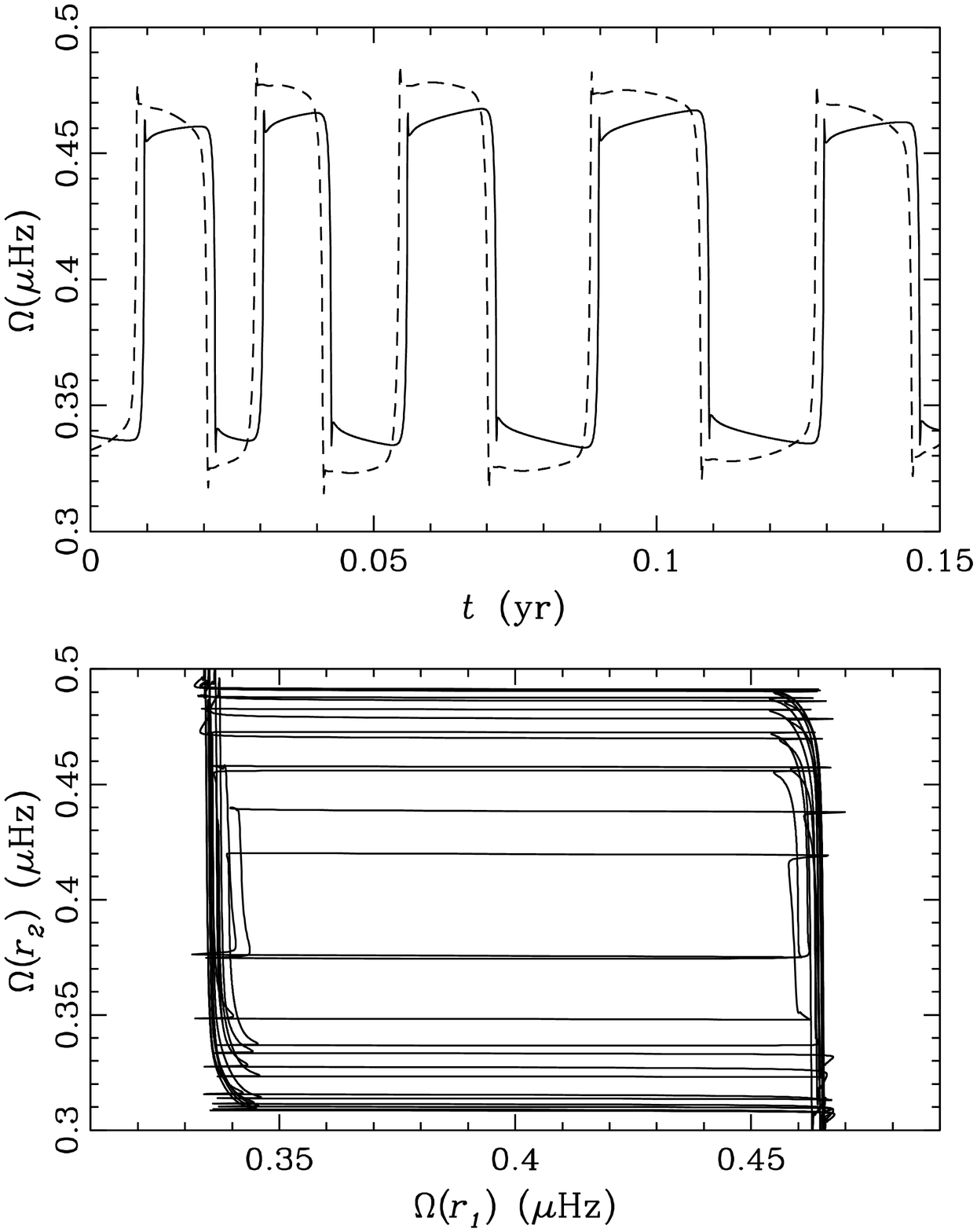}
}
\caption{Behavior of the shear layer as a function of the diminishing viscosity
for a $1.3\,M_\odot$ model with a turbulent viscosity given by
Eq.~(\ref{adhocvis}). From left to right, $N=8$, $N=7.9$ and $N=7.85$.
({\em top}) Time evolution of the rotation velocity at two different depths.
({\em Full line}) $\Omega(r_1)$ ({\em dashed line}) $\Omega(r_2)$
with $r_1=r_{cz}-0.003H_P$ and $r_2=r_{cz}-0.015H_P$.
({\em bottom}) Phase space diagram.
\label{chaos}}
\end{figure*}

\subsection{Wave Mean-Flow Interaction and the SLO \label{sec:SLO}}

It is now well established that the dissipation 
of internal gravity waves in a differentially rotating region leads 
to an increase in the local differential rotation.
In stellar models, this leads to the formation of
a narrow ($\sim$ 1-2\% in radius) doubled peak oscillating
shear layer adjacent
to the convection zone where they are produced
(Gough \& McIntyre 1998, Ringot 1998, Kumar, Talon, \& Zahn 1999).
We shall refer to this process as ``shear layer oscillation'' or SLO.

In a simple two waves model, Kim \& MacGregor (2001) examined
the behavior of that layer, which
depends on the ratio between viscosity and wave flux.
If the viscosity is large, a stationary solution may be found.
There then exists a bifurcation to an oscillatory behavior 
of the shear as viscosity
is reduced. Further reduction leads to the appearance of
chaos. The determination of the stellar viscosity is thus of
key importance for the structure of this layer.

If only radiative viscosity is considered, one observes 
the formation of a very steep and narrow
shear layer, and chaotic behavior is expected. However,
as the local shear increases, it may lead to the appearance 
of a shear instability, which will enhance the local
viscosity compared to its microscopic value. This process actually 
self-regulates the wave-mean flow interaction. Indeed, a
larger wave flux leads to a larger differential rotation which in
turns acts as to increase the local viscosity. 
  
Several calculations have been performed, using various prescriptions for the
turbulent viscosity. The first case-study should consist of using simply the radiative 
viscosity.
It has the major advantage that it can be derived from first principles only.
However, this viscosity is so low compared to the wave flux that it rapidly leads to a slow
layer that is stopped and even begins to rotate backward. To prevent that, we will
consider a ``turbulent'' viscosity proportional to the radiative viscosity 
\beq
\nu=10^N \nu_{\rm rad}. \label{adhocvis}
\eeq 
A regular oscillating layer is formed for $N=8$ (a similar form was used by 
TKZ). For a slightly larger ($N=9$) viscosity, we no longer obtain an oscillation.
This corresponds to the stationary solution found by Kim \& MacGregor (2003). Here, we do not
find a stationary profile due to the use of different boundary conditions (Neumann rather than
Dirichlet); our solution slowly decays to a state of solid body rotation.
For smaller ($N=7.9$ and $N=7.85$) viscosities, there is a transition to
chaos. These results generalize the Kim \& MacGregor (2001) results to the case where
a complete wave spectrum is considered. Period doubling is not readily
identified in this simulation.
Figure~\ref{chaos} summarizes our findings.

The turbulent viscosities used so far are not realistic. They vary only slowly with depth, and
remain large even far from regions where there exists a physical mechanism to produce this
turbulence. We must thus find a reliable prescription in order to estimate the magnitude of 
turbulence on physical grounds.

The structure of the shear region points to the shear instability as an important source of
turbulence. Generally, its magnitude depends on the local shear rate and on the efficiency
of buoyancy which acts as a restoring force. When taking into account radiative losses 
(Townsend 1958, Dudis 1974, Ligni\`eres et al. 1999) and the effect of
horizontal turbulence on the stabilizing effect of mean molecular weight gradients
(Talon \& Zahn 1997), the viscous turbulence 
may be approximated by
\beq
\nu_v = \frac{8}{5} \frac {Ri_{\rm crit}  (r \diff
 \Omega/\diff r)^2}{N^{2}_{T}/(K+D_h)+N^{2}_{\mu}/D_h}.
 \label{maeder}
\eeq

\begin{figure}[t]
\centering{
\includegraphics[width=6.5cm,angle=270]{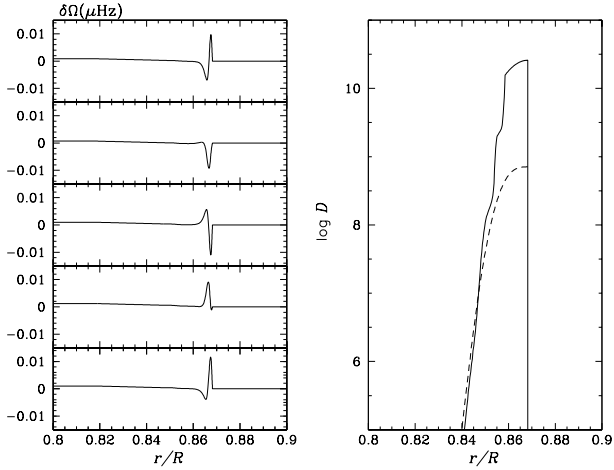}
}
\caption{({\em left}) Shear layer evolution (successive profiles are separated by
1~year) and ({\em right}) average turbulent
viscosity (as described in \S~\ref{sec:SLO}). The dashed line corresponds to an analytical fit
(\cf Eq.~\ref{nuwaves}).  
\label{shear}}
\end{figure}

This formulation is based on the local shear rate. In a rapidly varying profile,
such as the one we get in the shear layer, it leads to 
large gradients in the turbulent
diffusion coefficient, and is probably not realistic. Indeed, vertical turbulence
can develop over vertical scales similar to the pressure scale height 
which is greater than the scale of the variations of the shear layer 
(less than $1$\% in radius, see \eg Fig.~\ref{shear}, while in low
mass main sequence stars the pressure scale height just below 
the convection zone ranges from about $\sim\,10$\% in radius in 
a $1\,M_\odot$ star to $\sim\,3$\% in a $1.4\,M_\odot$ star). 
In practice, we thus used a convolution of the local
shear turbulence given by Eq.~(\ref{maeder}) with a Gaussian of width $0.2~H_P$ ($H_P$ is the
pressure scale height) in order to
obtain a more realistic turbulent coefficient\footnote{The Gaussian we use here
is quite thin. 
Far from a sharp shear layer, the convolution we are using would not alter the turbulent
viscosity from that obtained with Eq.~(\ref{maeder}).}. Turbulence is also averaged
over a complete oscillation cycle.

\begin{figure}[t]
\includegraphics[width=8cm]{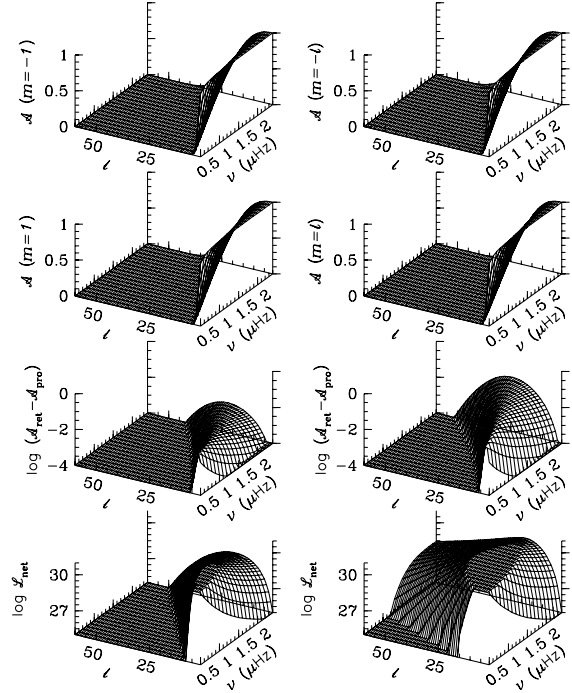}
\caption{Wave characteristics below the shear layer for a differential rotation
of $\delta \Omega = 0.1~\mu {\rm Hz}$ over $5\% $ in radius in a 1.3\,$M_\odot$ star.
{\em First and second row}: Wave amplitude for $m=\pm 1$ and $m=\pm \ell$.
{\em Third row}: Amplitude difference between retrograde and prograde waves of
the same azimuthal number.
{\em Fourth row}: Net luminosity for a given azimuthal number.
\label{lnet}}
\end{figure}

In this framework, the magnitude of the shear layer is self-regulated. An increase in the
wave flux leads to an increased differential rotation and thus, an increase of the shear
turbulence. 

Calculations 
are shown here for a $1.3\,M_\odot$, $Z=0.02$ model. Figure~\ref{shear} shows the
evolution of the shear layer as well as the self-consistent turbulent viscosity induced in
the model. In this model, thermal diffusion is rather large (see Talon \& Charbonnel 2003).
This explains the large value for the turbulent coefficient, and also its extreme thinness.
In lower mass stars, while just below the convection zone 
the viscosity is
large enough to lead to a regular oscillating layer, it then decreases rapidly
in the interior, leading to a complex layer structure that has a chaotic behavior.
However, this has no particular impact on long term momentum extraction.

This turbulence is generated by wave momentum deposition due to radiative damping,
and is related to turbulence produced by large amplitude waves where they are dissipated
(Press 1981, Canuto 2002, Young et al. 2003). It
is quite different in essence from the turbulent diffusion generally associated 
with waves,
either due to wave breaking (Press 1981, Garc\'\i a L\'opez \& Spruit 1991) or
to irreversible second order motions (Press 1981, Schatzman 1993, Montalb\'an 1994). 
This issue will be discussed further in \S~\ref{evochem}.

\subsection{Momentum Deposition Beyond the Shear Layer \label{sec:filter}}

The superficial shear layer acts as a filter on internal gravity waves
(Gough \& McIntyre 1998). Indeed,
as internal waves travel across first a ``rapid'' and then a ``slow'' layer,
prograde and retrograde waves respectively are preferentially
damped. However, some of the
power remains in the lowest order waves, and over long time-scales
(of the order of evolutionary time-scales), momentum redistribution 
below the shear layer cannot be neglected (TKZ). 
Figure~\ref{lnet} illustrates wave characteristics below the shear layer
for a given differential rotation, with the convection zone rotating slower than the
radiative zone.
The two top rows compare the amplitude of retrograde and prograde waves. As shown in the
third panel, the retrograde waves are somewhat less damped then their prograde
counterparts, and this leads to a net luminosity of momentum, illustrated in the bottom
panel. As explained in 
TKZ this is because the underlying differential rotation
produces a prograde layer which, on average, is always larger than the retrograde layer.

\begin{figure}[t]
\centering {
\includegraphics[width=8cm]{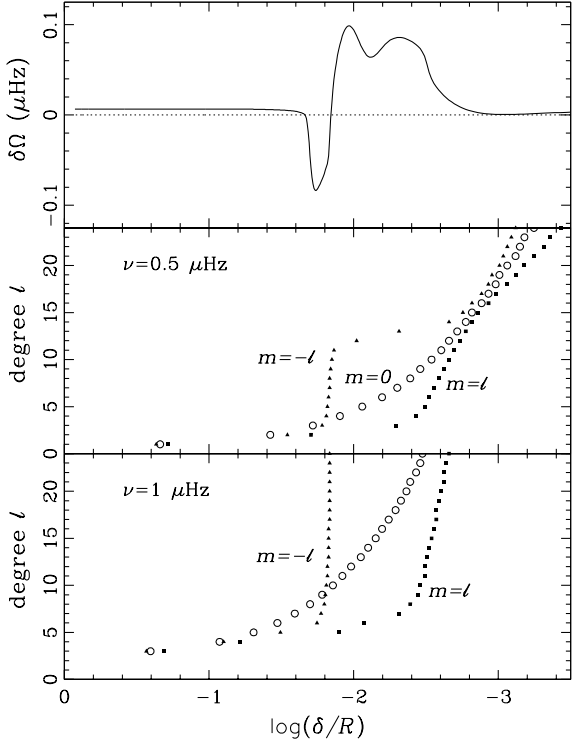}
}
\caption{
({\em top}) Instantaneous rotation profile
({\em middle, bottom}) Instantaneous damping length (corresponding to a reduction of wave
amplitude $A=A_0/e$)
for wave frequencies of
$\nu=0.5$ and $\nu=1~\mu{\rm Hz}$ in a 1.3\,$M_\odot$ star. The $m=0$ mode (open dots)
is not affected by
rotation, the $m=\pm \ell$ (squares and triangles) are the most affected.
Below the shear layer, the radiative region is rotating more rapidly; it is the retrograde
modes of low order that penetrate deeper, leading to a net deposition of negative momentum in
the interior. The low ($\ell =1$ and $2$) modes have to travel back and forth a few times
before they are damped, and are not shown in the bottom graph. 
\label{dampinglength}}
\end{figure}

It is the low frequency, low degree waves that give the largest contribution
to momentum redistribution in the interior; high degree waves are damped closer to the
convection zone (as damping $\propto \ell \lp \ell+1 \rp^{3/2}$) and thus have an
intrinsically small amplitude below the shear layer
while high frequency
waves experience less differential damping ($\propto 1/\sigma(r)^4 = 1/\lc \omega - m
\lp \Omega(r)-\Omega _{\rm cz} \rp \rc ^4$), crucial to produce a net momentum deposition
(see Fig.~\ref{dampinglength}).
See \S~\ref{modelang} for details on damping.

In order to built evolutionary stellar models, it is not possible to follow in
detail the behavior of the oscillating shear layer, which occurs on time-scales
of years or tens of years. However, this shear layer is crucial in filtering
the low order waves that travel through it. The net wave flux across the shear
layer depends of the asymmetry of the filter and thus, on the difference between
the rotation velocity below the shear layer and that of the convection zone.
Let us note however that the thickness of the oscillating layer is independent of
this asymmetry.
Last but not least, wave momentum
transport does not have a diffusive behavior and as such must not be treated
merely as a turbulent diffusion process.

Figure~\ref{dampinglength} illustrates differential damping in
a shear region, with both a ``rapid'' and a
``slow'' layer. The $m=0$ mode is not affected by the differential rotation\footnote{This 
mode does not carry angular momentum.},
and, for a given frequency, 
the damping length $\delta$ is inversely proportional to $\ell (\ell+1)$.
The ``rapid'' layer increases the damping of prograde modes, and reduces it for
the retrograde modes. The reverse is obtained from the ``slow'' layer. 

\begin{figure}[t]
\centering {
\includegraphics[width=8cm]{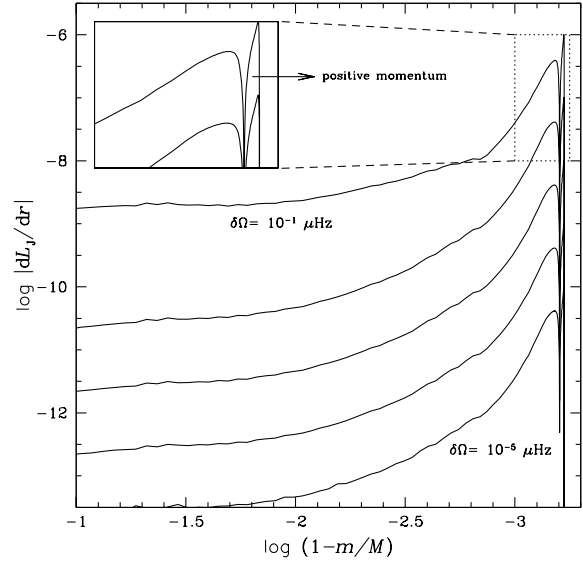}
}
\caption{Local deposition of angular momentum for various values of
differential rotation. The rotation profile rises linearly between the base of
the
convection zone and $(1-m/M)=-2.85$ by a value of $\delta \Omega$, which varies
logarithmically from $10^{-5}$ to $10^{-1}$~$\mu{\rm Hz}$. This is illustrated
for a $1.3\,M_\odot$ model. For moderate amounts of differential rotation
(below $\sim~10^{-2}~\mu{\rm Hz}$), the net momentum flux at a given point
is a linear function of differential rotation.
\label{dampingvsdomega}}
\end{figure}

Figure~\ref{dampingvsdomega} shows the local momentum deposition for various values of
differential rotation. As far as differential rotation is not too large, momentum
deposition varies linearly with $\delta \Omega$. Actual calculations of the evolution of
the distribution of angular momentum show that, for realistic values of braking
(according to Kawaler 1988), the star remains in this linear regime.

\subsection{Energy Considerations}

It has been shown elsewhere (Kumar \& Quataert 1997, Zahn et al. 1997, 
TKZ) that gravity waves can carry enough angular momentum to slow 
down the radiative zone of low mass stars on time-scales of order $10^7$~years.
One may wonder however if the deposition of energy by waves may have an impact on the stellar
structure. Let us first begin by comparing the amount of energy contained in gravity waves
with other quantities. This will be done here for a ZAMS $1\,M_\odot$ model.

The total energy luminosity in waves is of order $10^{29}~{\rm ergs\,s^{-1}}$, while the total
luminosity at the base of the convection zone is $2.5 \times 10^{33}~{\rm ergs\,s^{-1}}$, of which a 
large part is convective (see Fig.~\ref{excitation}). The wave luminosity thus represents 
about $0.01$\% of the convective luminosity. 

Let us next look at the amount of kinetic energy that is stored in rotation $K_\Omega$. We have 
$K_\Omega = \frac{1}{2} I \Omega^2$ with $I$ the moment of inertia of the star. For the whole star,
it is of order $10^{54}~{\rm cm^2\,g}$. For an initial rotation velocity of $100~{\rm km\,s^{-1}}$,
this leads to an energy of $10^{46}~{\rm ergs}$. As the star is spun down, this energy
must be dissipated and will be added to the thermal energy.
However, the total rotation energy is two order of magnitude lower than the thermal
content and so, the impact will be negligible on the stellar structure.

\section{A Model for Angular Momentum Evolution by Waves \label{modelang}}

Gravity waves lead to two different features that must be incorporated in stellar
evolution codes: 
\begin{itemize}
\item They produce a shear layer, that generates turbulence close to the bottom
of the convection zone;
\item They deposit negative (positive) momentum throughout the radiative interior when
the convection zone rotates slower (faster) than the radiative zone.
\end{itemize}

The turbulent region is always present, and the shear layer oscillation is (almost)
independent of the presence or absence of differential rotation below.
The magnitude of the turbulence and the size of the turbulent region are self-regulated
and depend on the wave flux. For a given wave-excitation model, and a given model for
the turbulent diffusion, we obtain a localized turbulent region.

\begin{figure*}[t]
\centering {
\includegraphics[width=8cm,angle=270]{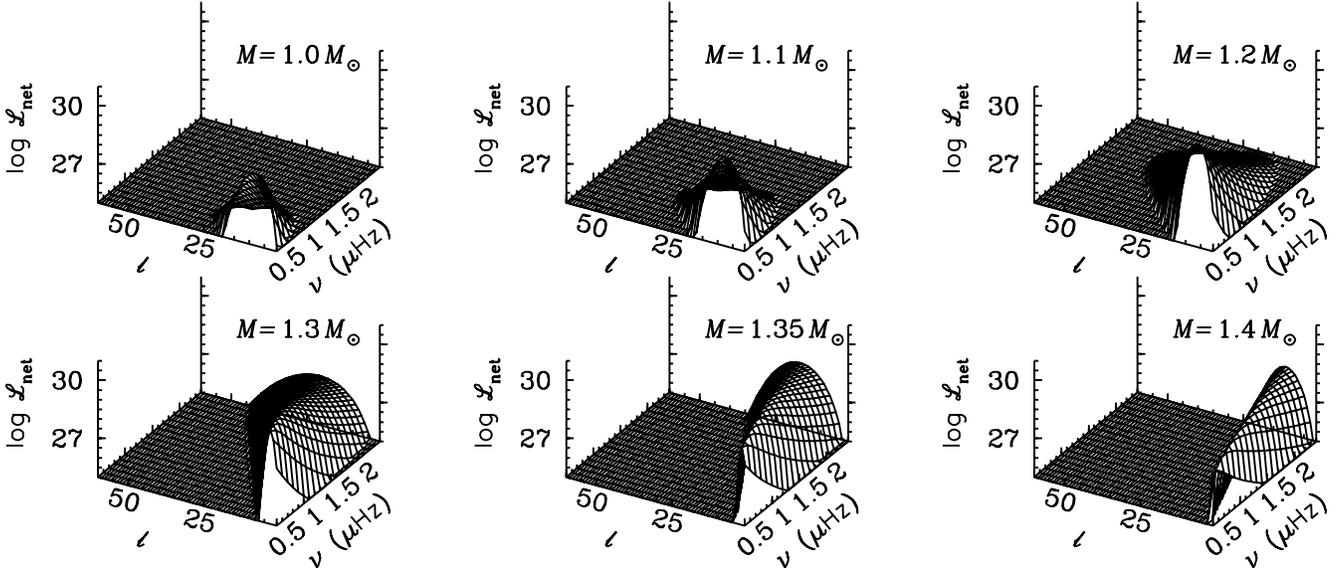}
}
\caption{Filtered momentum luminosity spectrum for the $m=\ell$ mode (that is ${\cal L}_{\rm net}
= {\cal L}_{m=-\ell} - {\cal L}_{m=\ell}$) below the shear layer 
for various masses. Differential rotation is $\delta \Omega = 10^{-4}~\mu {\rm Hz}$ 
over $5\% $ in radius. 
\label{lnetmass}}
\end{figure*}

The net momentum luminosity spectrum below the shear layer for various masses is shown
in Fig.~\ref{lnetmass}. To obtain the net momentum deposition, one must follow the local
momentum luminosity
\beq
{\cal L}_J(r) = \sum_{\sigma, \ell, m} {{\cal L}_J}_{\ell, m} \lp r_{\rm shear~layer}\rp
\exp \lc -\tau(r, \sigma, \ell) \rc \label{locmomlum}
\eeq
and where the local damping rate takes into account the mean molecular weight
stratification
\beq
\tau(r, \sigma, \ell) = [\ell(\ell+1)]^{3\over2} \int_r^{r_c} 
\lp K + \nu_t \rp \; {N N_T^2 \over
\sigma^4}  \left({N^2 \over N^2 - \sigma^2}\right)^{1 \over 2} {\diff r
\over r^3} \label{optdepth}
\eeq
where $N^2 = N_T^2 + N_{\mu}^2$ is the total Brunt-V\"ais\"al\"a frequency,
$N_T^2$ is its thermal part and $N_{\mu}^2$ is due to the
mean molecular weight stratification (\cf Zahn et al. 1997).
$\sigma$ is the local, Doppler shifted frequency
\beq
\sigma(r) = \omega - m
\lp \Omega(r)-\Omega _{\rm cz} \rp \label{sigma}
\eeq
and $\omega$ is the wave frequency in the reference frame of the convection
zone.
Only those waves with a sufficient amplitude after filtering have to be traced.

When meridional circulation, turbulence and waves are taken into account, the evolution
of angular momentum thus follows
\beqan
\rho \dtt \lc r^2 {\Omega}\rc &= &
\frac{1}{5 r^2} \drr \lc \rho r^4 \Omega U \rc 
+ \frac{1}{ r^2} \drr \lc \rho (\nu_t + \nu_{\rm waves}) r^4 \dr{\Omega} \rc 
\nonumber \\
&& - \frac{3}{8\pi} \frac{1}{r^2} \drr{{\cal L}_J(r)},
\label{ev_omega}
\eeqan
where $\rho$ is the density, $U$ the radial meridional circulation velocity,  
$\nu_t$ the turbulent viscosity due to differential rotation away from the shear layer,
and $\nu_{\rm waves}$ the diffusion coefficient associated with wave-induced turbulence
(see Eq.~\ref{nuwaves} in \S~7.1).
Horizontal averaging has been performed for this equation, and meridional circulation is
considered only at first order (see \S~\ref{Description}).

\section{Evolution of Chemicals \label{evochem}}

In most hydrodynamical models, 
chemicals evolve under the action of meridional circulation,
turbulence and atomic diffusion. All these topics are discussed
elsewhere (see \eg Talon 2004 and references therein for a discussion of
rotational mixing, and Turcotte et al. 1998 and references therein 
for a review of microscopic diffusion processes).
However, when including gravity waves in momentum transport, one must also
include their direct contribution to the transport of chemical
species\footnote{This adds to the indirect effect described above, that is the modification
of the rotation profile which changes rotational mixing.}.
This subject has received attention from several authors and we will here 
describe the main mechanisms involved.

The first process we shall discuss here is the turbulence generated in 
the shear layer by the shear instability. In a first step, energy is transfered 
by waves from the convection zone to the shear layer and stored into 
differential rotation, which can then be converted to turbulence by the shear
instability. This view is similar to the idea of Canuto (2002) that gravity waves
act as a source term in the equation that describes turbulence. This is
also similar to the mixing described by Young et al.
(2003). In our framework, turbulent diffusion is evaluated by taking the average 
of Eq.~(\ref{maeder}) over several SLO.

Weak mixing can also be induced by second order mass-transport
effects in a diffusive medium (Press 1981, Schatzman 1993, Montalb\'an 1994).
A diffusion coefficient can then be associated with wave dissipation
by combining the average 
wave velocity and the average damping length of waves.
If the concomitant
transport of angular momentum is taken into account, the damping length of waves
is reduced and the effect is to somewhat lower the size of the region
over which this process is efficient. 
The thickness of the shear layer should, in many instances replace the damping 
length calculated in the case of solid body rotation.
Differential rotation thus reduces the extent of this effect; its magnitude is
at maximum of the same order as turbulence induced by the SLO. Considering the 
uncertainties in waves fluxes as well as on the mixing induced by those motions 
we suggest to neglect this effect altogether.

Finally, in certain circumstances, wave amplitude can rise to the point of
becoming non-linear and inducing shear-mixing (Press 1981, Garc\'{\i}a L\'opez 
\& Spruit 1991). 
The non-linearity is given by
\beq
\epsilon \equiv k_h \xi _h = \frac{k_h u_h}{\sigma}
\eeq
(Press 1981).
The wave becomes non-linear when this parameter is of order 1 or more.
In terms of energy luminosity, it is given by
\beq
\epsilon = \frac{k_h}{\sigma} \lp \frac{{\cal L}_J \lp r,\sigma,\ell \rp}{4 \pi \rho r^2} \frac{k_h}
{\lp N^2 - \sigma ^2 \rp ^{1/2}} \rp ^{1/2}.
\eeq
As estimated by Press (1981), some waves are slightly non-linear just below
the convection zone; this non-linearity remains only over one damping length. 
However, since this region is already highly turbulent,
this small non-linearity does not greatly modifies wave damping. Close to
wave turning points (where $\sigma \rightarrow 0$) however, non-linearity may again
become important. This will affect mostly low degree, low frequency waves by
increasing their damping in that region. However, since radiative damping is already
strong there, it will not greatly modify wave amplitudes. However, in some cases it
could increase the amount of mixing of chemicals. Garc\'{\i}a L\'opez 
\& Spruit (1991) showed this has a significant effect only in a small part
of the HR diagram (namely in stars close to the Li dip for a specific choice of
the mixing length parameter). We will discuss the relevance of this effect
in a future paper, where models of various masses will be implemented.

\section{Waves in a Static Model}

\begin{figure}[t]
\centering{
\includegraphics[width=6.5cm,angle=270]{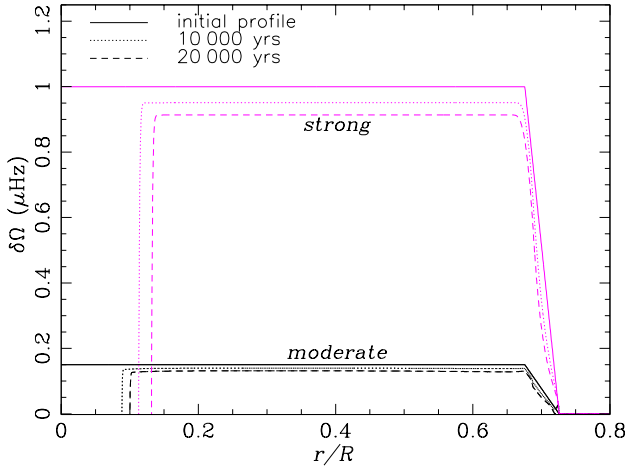}
}
\caption{Evolution of the rotation profile in the case of very strong 
and moderate initial differential rotation in a $1~M_\odot$ star.
\label{comprdiff}}
\end{figure}

Before we apply our filtered wave model to an evolutionary calculation, 
let us begin  by discussing some more the results obtained by 
TKZ in the case of a static model. Those 
are based on a rather small differential rotation, and use a given wave spectrum.
It is worth verifying how sensitive are those results to other initial profiles
or different distributions of energy.
In the models presented in this section, we show results obtained for small time steps
(of 1 year, as in TKZ),
following the details of the shear layer oscillation. The only other process for momentum
transport that is considered here is shear turbulence, and no surface braking is
applied\footnote{However, the initial rotation profile has a surface convection
zone rotating slower than the radiative interior.}. Contrary to the TKZ 
study, here turbulence is directly related to the local shear rate (\cf Eq.~\ref{maeder}). 

\subsection{Strong Initial Differential Rotation}

Let us discuss some more results presented in \S~\ref{sec:filter}. In
Fig.~\ref{comprdiff}, we compare the evolution of the rotation profile in cases
of very strong and moderate\footnote{This is the same as was used by TKZ.} 
differential rotation. Momentum extraction from the
core is clearly visible in both case. The core's slow rotation slowly propagates
toward
the surface. It does so with a velocity that gets smaller as the front 
progresses to
a region where the local angular momentum ($\propto r^2$) is larger. 
Differential rotation at the core boundary
remains larger than in TKZ because the local turbulence $\nu_t$ is smaller\footnote{In
TKZ, the local turbulence was fixed (as in Eq.~\ref{adhocvis}) 
and did not depend on the local shear rate.}. The
important point to note is that in the presence of a large differential
rotation, the local frequency of retrograde waves becomes very large and their
damping is largely reduced in the inner regions. However, once the core has been
spun down which is easily done since it contains very little momentum, a
``slow'' front can propagate toward the surface. 

\begin{figure}[t]
\centering{
\includegraphics[width=6.5cm,angle=270]{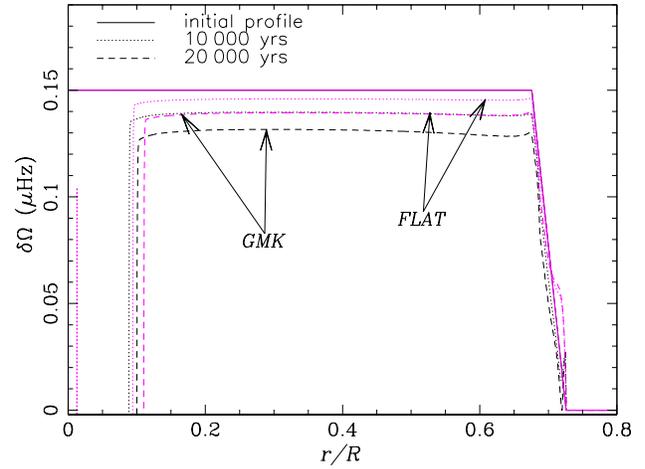}
}
\caption{Evolution of the rotation profile with moderate initial differential rotation for
a wave spectrum derived from GMK spectrum (\cf Eq.~\ref{gold}) and
for a flat spectrum with the same total energy flux. Calculations are made in a $1~M_\odot$ star
\label{compflux}}
\end{figure}

\subsection{Wave Spectrum}

Another delicate point is the form of the wave spectrum. As already mentioned
(\S~\ref{sec:waveexcitation}), the exact spectrum 
produced by turbulence in the convection zone is not well constrained, and various
analytical and numerical studies lead to different prescriptions. It is thus important to
understand how modifying the spectrum influences the results.
Figure~\ref{compflux} illustrates this comparison. The flat spectrum produces a front that
migrates somewhat faster, but deposits less momentum in 
the intermediate region. In that case, the shear
layer does not oscillate, but still produces a filter that preferentially damps 
prograde waves. While the exact spectrum changes the results quantitatively,
qualitatively global effects are similar.

\section{Waves in an Evolutionary $1.2\,M_\odot$ Model}

\subsection{Code Description \label{Description}}

\noindent{\em General inputs for stellar physics}\\
Stellar models are computed with the stellar evolution code STAREVOL 
(Forestini 1991; Siess et al. 1997, 2000; Palacios et al. 2003). 
Our equation of state follows the Pols et
al. (1995) formalism. Thermodynamical features of each plasma component
(ions, electrons, photons and $\mathrm{H}_{2}$)
are obtained by minimizing the Helmholtz free energy that includes separately
non-ideal effects, and allows to treat ionization analytically, leading to smooth
profiles for thermodynamical quantities. Radiative opacities are taken from 
Alexander \& Fergusson (1994) below 8000~K and from Iglesias \& Rogers (1996) at 
higher temperatures. Our nuclear reaction network follows 
53 species (from ${}^{1}\mathrm{H}$ to
${}^{37}\mathrm{Cl}$) through 180 reactions. Nuclear reaction rates
have been updated with the NACRE compilation (Angulo et al. 1999).
Convection is treated according to the mixing length formalism with
$\alpha_p = l/H_p=1.75$. No overshooting is considered.\\

\noindent{\em Rotation}\\
We follow the evolution of the rotation profile from the zero age main sequence on,
assuming initial solid body rotation. The surface rotation velocity on the ZAMS is 
taken equal to $50~{\rm km\,s^{-1}}$. Surface spin-down follows
Kawaler (1988) including saturation at $\Omega_{\rm sat} = 1.5 \times 10^{-5}~{\rm rad\,s^{-1}}$ 
(or $v \sim 10~{\rm km\,s^{-1}}$)
\begin{equation}
 \frac{{\rm d} J}{{\rm d}t} = \left\{
\begin{array}{l l }
-K_S \Omega^3 \lp {\displaystyle {R}/{R_\odot}} \rp ^{1/2} 
\lp {\displaystyle {M}/{M_\odot} }\rp ^{-1/2} & 
(\Omega \leq \Omega_{\rm sat}) \\
 & \\
-K_S \Omega \, {\Omega^2}_{\rm sat} \lp {\displaystyle {R}/{R_\odot}} \rp ^{1/2} 
\lp {\displaystyle {M}/{M_\odot} }\rp ^{-1/2}
& (\Omega > \Omega_{\rm sat}).  
 \end{array}   \right.
\end{equation}
The issue of saturation in the context of meridional circulation is discussed at
length in Palacios et~al.~(2003). \\

\noindent{\em IGW}\\
To incorporate IGW, we first calculate filtered luminosities 
(\cf Fig.~\ref{lnetmass}) for fixed differential rotations ($\delta \Omega=10^{-4},
~10^{-3},~10^{-2}~{\rm and}~10^{-1}~\mu {\rm Hz}$) in static ZAMS 
models\footnote{We showed in Talon \& Charbonnel (2003) 
that in the type of star we consider for this full evolutionary 
computation (i.e., Pop I main sequence star with a mass of 1.2M$_{\odot}$) 
the depth and structure of the convective envelope, 
and thus the wave characteristics, do not vary significantly 
over a main sequence lifetime. It is thus justified to perform the
first test with the spectrum characteristics of the ZAMS models}.
The wave luminosity below the surface convection zone is then
linearly interpolated from those tables for the actual differential rotation
just below the shear layer. The local momentum luminosity is obtained by calculating the damping
integral for each individual wave (\cf Eq.~\ref{optdepth}) 
and then summing over all waves
(\cf Eq.~\ref{locmomlum}). \\

\noindent{\em Meridional circulation} \\
Meridional circulation is treated as an advection process
for the transport of angular momentum, 
assuming strong horizontal turbulence and follows Zahn (1992).\\

\noindent{\em Turbulence and diffusion} \\
Turbulence is assumed to be strongly anisotropic, and everywhere we will assume
that turbulent diffusion is equal to the turbulent viscosity. 
The vertical component of the turbulent viscosity 
\beq
 D_v = \nu_v = \frac{8}{5} \frac {Ri_{\rm crit}  (r \diff
 \Omega/\diff r)^2}{N^{2}_{T}/(K+D_h)+N^{2}_{\mu}/D_h}
\label{Dv}
\eeq
takes into account the weakening effect of thermal diffusivity ($K$) on the thermal
stratification and of horizontal turbulence ($D_h$) on both the thermal and mean
molecular weight stratifications (Talon \& Zahn 1997).

\noindent Horizontal turbulence follows Zahn (1992)
\beq
D_h = \nu_h = \frac{r}{C_h}\left|\frac{1}{3 \rho r}\frac{\diff (\rho r^2 U)}{\diff
    r}-\frac{U}{2}\frac{\diff \ln r^2\Omega}{\diff \ln r}\right|.
\label{Dh}
\eeq
with $C_h=1$.
The evolution of momentum is then given by Eq.~(\ref{ev_omega}).\\

\noindent{\em Transport of chemicals} \\
For chemicals, the combination of meridional circulation and horizontal turbulence
results in a vertical effective diffusion
\beq
D_{\rm eff} = \frac{ \left| r U(r) \right|^2}{30\,D_h}
\eeq
(Chaboyer \& Zahn, 1992).

\noindent Atomic diffusion driven by gravitational settling 
and thermal gradients is included using the formulation of Paquette et~al.~(1986).

\noindent The evolution of an element $c_i$ then follows
\beq
\rho  \frac{{\rm d} {c_i}}{{\rm d} t}  =  \dot{c_i} +
\frac{1}{r^2}\frac{\partial}{\partial r}\left[r^2\rho  \left\{ U_{ip}{c_i}
  + 
   \left(D_{\rm eff}+D_{v}+D_{\rm waves}\right)\frac{\partial {c_i}}{\partial r} \right\} \right]
\label{diffelts2}
\eeq
with $c_i$ the nuclear production/destruction rate and $U_{ip}$ the microscopic
diffusion velocity with respect to protons. 

\noindent Wave induced turbulence is treated
explicitly through coefficient 
$D_{\rm waves}$ that we assume equal to the coefficient $\nu_{\rm waves}$
which enters Eq.~\ref{ev_omega}. 
To evaluate this coefficient, we 
perform a time average of the diffusion coefficient associated with the
local shear instability (Eq.~\ref{Dv}) over several
SLOs (see Fig.~\ref{shear}, and \S~\ref{sec:SLO}). For evolutionary calculations,
we use an analytical fit to this curve.
In this application to a $1.2~M_\odot$ star, we used
\beq
\log D_{\rm waves}= \lp \log K + 2 \log L -68 \rp \exp \lc \lp r_{\rm cz}-r \rp^3/\sigma \rc 
\label{nuwaves}
\eeq
with
\beq
\sigma = 0.0003 \lp 1 - r_{\rm cz} \rp.
\eeq
This fit is compared to the time-average in Fig.~\ref{shear}.
Let us remark that above $\sim 10^8~{\rm cm^2\,s^{-1}}$, mixing may be considered
``instantaneous'' compared with evolutionary time-scales and thus, our analytical fit is
aimed at reproducing the numerical turbulent viscosity below this value.

\subsection{Evolution of the Rotation Profile \label{sec:resultsangmom}}

\begin{figure}[t]
\centering {
\includegraphics[width=8cm]{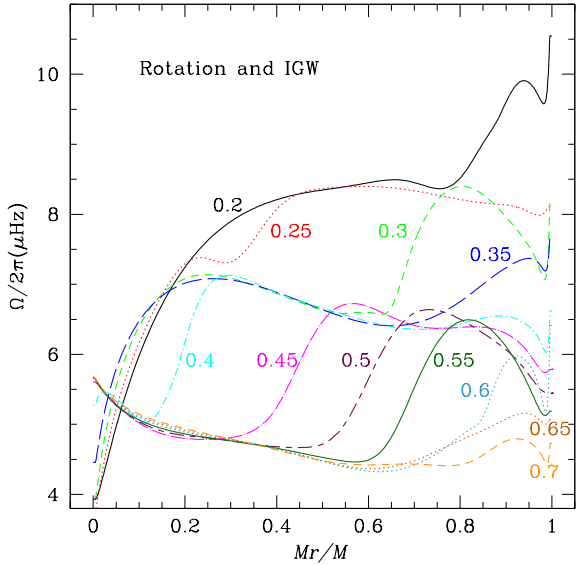}
}
\caption{Evolution of the rotation profile in a complete model where the transport 
of angular momentum is due to internal gravity waves, meridional circulation
and turbulence. The model shown is for a $1.2\,M_\odot$
$Z=0.02$, star with 
an initial rotation velocity of $50~{\rm km\,s^{-1}}$. 
The curves are labeled according to the corresponding ages 
in Gyr.
\label{evoomega}}
\end{figure}

\begin{figure}[t]
\centering {
\includegraphics[width=7cm,angle=270]{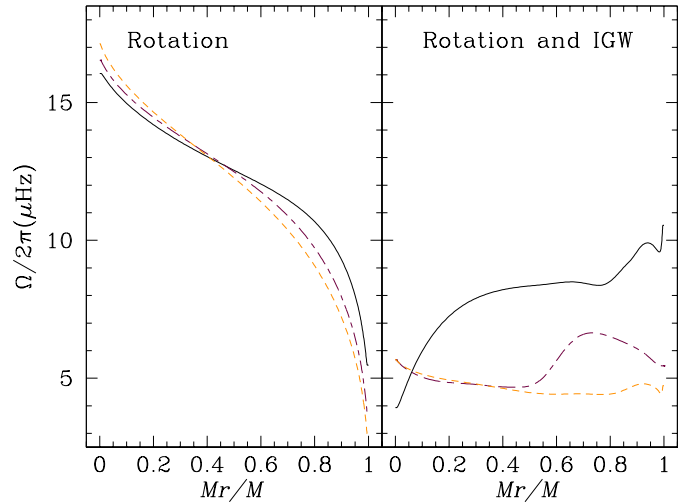}
}
\caption{Same as Fig.~\ref{evoomega} for 0.2, 0.5 and 0.7~Gyr in a case 
without ({\em left}) and with ({\em right}) internal gravity waves included.
The time sequence goes from the highest to the smallest surface velocity
\label{evoomegacompar}}
\end{figure}

Let us first concentrate on the evolution of the rotation profile when momentum
deposition by IGW is taken into account in conjunction with shear turbulence and meridional
circulation.
Here, the short time-scale SLO is present only as a filter; the magnitude of the wave
flux depends on the amount of differential rotation between the base of the
convection zone and a region just below the SLO (see \S~\ref{sec:filter}).
In quasi-solid body rotation (with the surface rotating slightly slower because of
braking), low degree waves penetrate all the way to the stellar
core, and are damped (and thus deposit their negative momentum) over the whole
radiative region. However, since the amount of angular momentum contained in the 
radiative core is minute, the local deposition of even a small amount of momentum 
is enough to spin it down
significantly. In a ``slow'' region, damping of retrograde waves increases
(\cf Eqs.~\ref{optdepth} and \ref{sigma}), and this
leads to the formation of a ``front'', which propagates from the core to the surface.
The propagation of a first front is seen in Fig.~\ref{evoomega} (curves at 0.2, 0.25, 0.3
and 0.35~Gyr).
A second front evolves from 0.4 to the 0.7~Gyr curves.

At the age of the Hyades, differential rotation is considerably reduced. 
In particular, it is of interest to compare calculations
including and not including waves. In the second case (left plot in Fig.~\ref{evoomegacompar}),
the amount of differential rotation at the age of the Hyades is very large. This is in 
in agreement with the results obtained in the solar case by Matias \& Zahn (1998) 
who performed calculations under the same hypothesis as here, and by Chaboyer et~al.~(1995) 
who approximated meridional circulation as a diffusive process.

The present complete model confirms the ability of gravity waves to efficiently 
extract angular momentum from the deep interior of solar-type stars.  
It shows how the momentum redistribution proceeds when the stars are 
spun down via magnetic torquing. 

Let us see now what are the consequences on meridional circulation and shear turbulence,
and on the Li depletion due to rotational mixing.

\subsection{Meridional Circulation Velocity and Diffusion Coefficients \label{sec:diffuscoefs}}

\begin{figure}[t]
\centering {
\includegraphics[width=8cm]{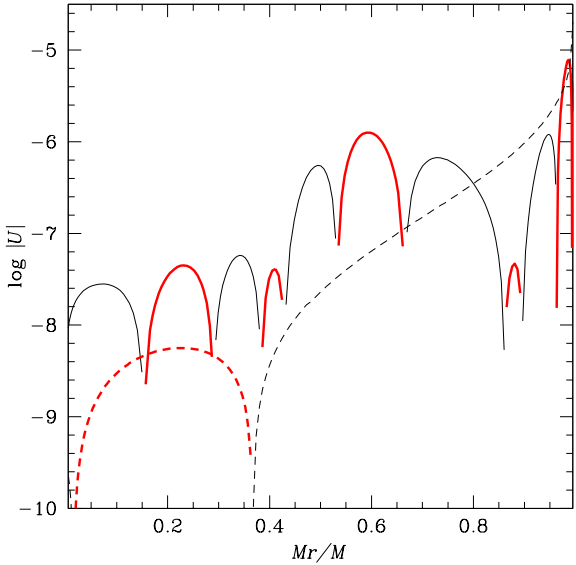}
}
\caption{Profiles of the vertical component of the meridional velocity 
in a $1.2\,M_\odot$ star at 0.5~Gyr in a simple rotating model (dotted line) 
and in a rotating model including gravity waves (full line). The bold 
lines indicate positive values of $U$.
\label{Ucirc}}
\end{figure}

Figure~\ref{Ucirc} presents profiles of the vertical component of the meridional
velocity $U$ at 0.5~Gyr in the $1.2~M_{\odot}$ models with and without gravity waves. 
When not including waves, there are two circulation loops. The meridional velocity
is negative in the external part of the radiative zone down to $\sim 0.3\,M_*$.
This meridian loop brings matter upward at the equator and down in the 
polar regions, in response to the extraction of angular momentum due to braking.
Deeper the circulation is positive (bold line), indicating an inward transport of 
angular momentum. 
When internal waves are taken into account, several loops of circulation appear,
with negative and positive loops (light and bold lines respectively) alternating. 

\begin{figure}[t]
\centering {
\includegraphics[width=7cm,angle=270]{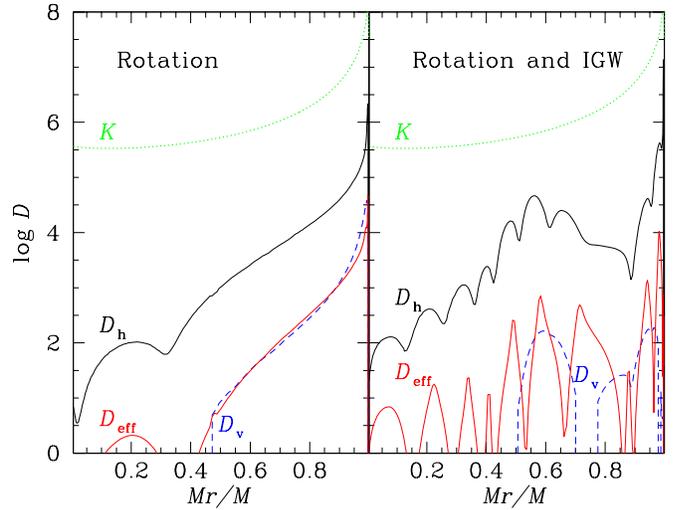}
}
\caption{Comparison of the transport coefficients in a $1.2\,M_\odot$ star at 
0.5~Gyr in a simple rotating model ({\em left}) and in a rotating model including IGW
({\em right}). Coefficients are identified on the figure. $D_v$ is the dashed line and
$D_{\rm eff}$ a full line.
\label{coefdiff}}
\end{figure}

\begin{figure}[t]
\centering {
\includegraphics[width=8cm]{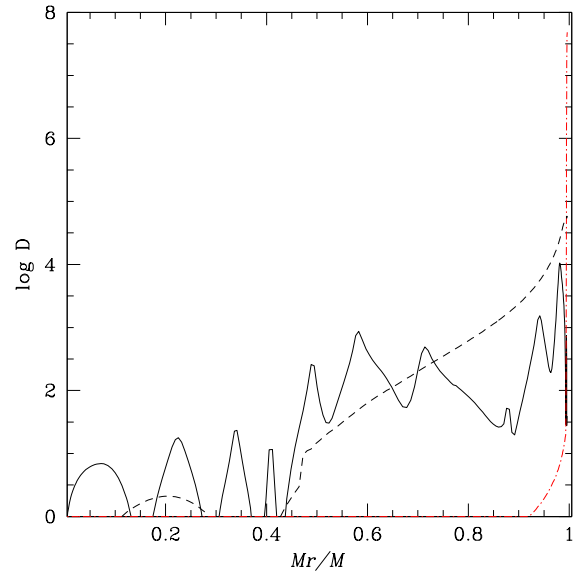}
}
\caption{Total diffusion coefficient in a $1.2\,M_\odot$ star at 
0.5~Gyr in a simple rotating model (dotted line) and in a rotating model 
including IGW (full line). 
The dashed-dotted line gives $D_{\rm waves}$.
\label{Dtot}}
\end{figure}

These differences reflect directly in the transport coefficients 
which are shown at the same age (0.5~Gyr) in Figs.~\ref{coefdiff} and \ref{Dtot}. 
Details in the profiles vary 
with time but this is a typical illustration.  
Strong variations (\ie peaks and gaps) in the profiles of 
both effective and total diffusion coefficients 
reflect the presence of several circulation loops, each drop corresponding to 
an inversion of the flow direction. When gravity waves are 
included, the amplitude of these coefficients is reduced and vertical turbulence ($D_v$)  
is less developed than in the pure rotating models because of the decrease of
the overall differential rotation. 

In Fig.~\ref{Dtot}, the wave-induced turbulence 
($D_{\rm waves}$) used in the present simulation with IGW 
and given by Eq.~\ref{nuwaves} is also shown. 
As can be seen, this coefficient drops very rapidly below the convection envelope
and is much smaller than the total diffusion coefficient coming from rotation
except very close from the convection region.

\subsection{Evolution of Chemicals \label{sec:resultschem}}

Our goal in the present paper is not to make detailed comparison of the 
model predictions with all the observable constraints that come from 
abundance anomalies in stars. This will be done in forthcoming studies.
We wish however to illustrate briefly the behavior of $^4$He 
and $^7$Li in our $1.2~M_{\odot}$ rotating model including gravity waves.
The case of $^4$He is related to possible impact on the overall evolution
and lifetimes, and in addition it illustrates the interaction with 
atomic diffusion. On the other hand $^7$Li is a fragile element which
helps to probe the status of the external stellar layers. 

\begin{figure}[t]
\centering {
\includegraphics[width=8cm]{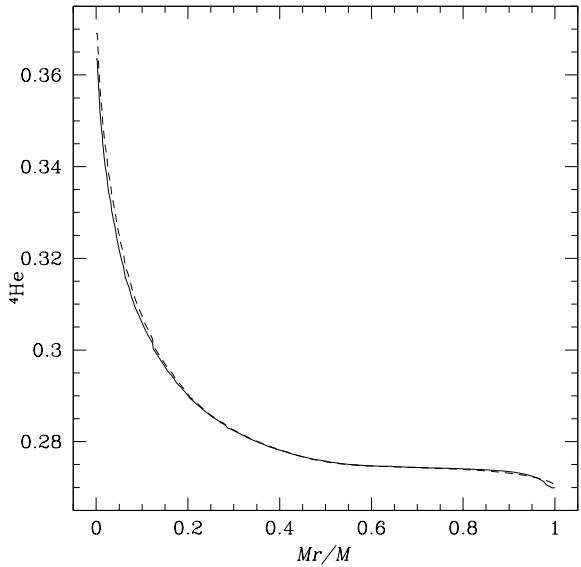}
}
\caption{$^4$He profile at the age of the Hyades in the complete model (full line)
and in the model without the IGW (dashed line)
for a $1.2\,M_\odot$ star.
\label{4He}}
\end{figure}

The $^4$He profile at the age of the Hyades is shown in Fig.~\ref{4He}
for both the complete model and the model without IGW.
One sees there two effects. First, at this age, the model with IGW
is slightly less evolved (\ie it has a lower central $^4$He content).
Second, its $^4$He surface abundance is slightly lower and the $^4$He
gradient just below the convection envelope is slightly steeper.
This is due to the lower amplitude of the total diffusion coefficient
in the models including waves, which allows $^4$He to settle more
under the effect of atomic diffusion.
One sees here that $D_{\rm waves}$ has a negligible effect on 
the  $^4$He behavior and does not compensate for the decrease in the 
total diffusion coefficient in the external part of the radiative zone.

\begin{figure}[t]
\centering {
\includegraphics[width=8cm]{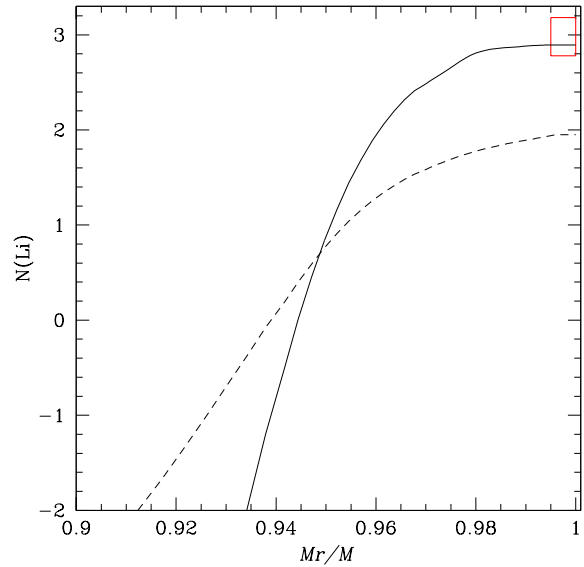}
}
\caption{$^7$Li profile at the age of the Hyades in the complete model (full line)
and in the model without the IGW (dashed line) 
for a $1.2\,M_\odot$ star. 
The box indicates the observed 
Li abundances in Hyades stars with an effective temperature ($\pm$ 100~K) 
corresponding to that of the model at 0.7~Gyr (data by Boesgaard \& King 2002)
\label{7Li}}
\end{figure}

The best observational constraints available to test our predictions 
are the Li data in open clusters. As discussed in \S~1, models  
in which the transport of angular momentum is carried out only by 
meridional circulation and turbulence fail to reproduce the rise of 
the Li abundance on the right side of the Li dip. This is confirmed
in the present model without IGW which lies in this region and has an effective
temperature of $T_{\rm eff} = 6220~{\rm K}$ 
at the age of the Hyades. As can be seen in Fig.~\ref{7Li}, the $^7$Li 
surface abundance in this model (dotted line) is $\sim$ one order of 
magnitude smaller than the Li value in stars of $6220 \pm 100~{\rm K}$ 
in the Hyades and which is indicated by the box. 
For the model including IGW (full line), the magnitude of both meridional circulation
and turbulence is reduced. Consequently $^7$Li is less destroyed and 
our prediction accounts nicely for the data.
Note that the surface lithium decrease in these stars is not 
dominated by atomic diffusion,
but still by the (reduced, compared to the case without IGW) rotational transport 
of this element down to regions where it is nuclearly destroyed.
Again, the effect of $D_{\rm waves}$ is negligible.

In a forthcoming paper, we will present models of Pop~I stars 
with various initial masses and initial rotation rates and compare in details 
the Li predictions on the red side of the Li dip with observations
in open cluster and field stars.

\section{Discussion and Conclusions}

In this paper, we examined how it is possible to include all 
the effects of internal gravity
waves into evolutionary calculations. The main challenge is that IGW tend to produce a thin
shear layer that oscillates with a very small time-scale. Talon et al. (2002) showed that this
shear layer oscillation (or SLO) acts as a filter for waves, being more transparent to waves
that will reduce the differential rotation between the convection and the radiation zones. We
verified that this filter is linear in $\delta \Omega$ and as such, details of the SLO need not
be considered for variations occurring over long time-scales. 

We also presented the first evolution model that includes the hydrodynamical processes 
induced by rotation and internal gravity waves. We focused on a $1.2~M_{\odot}$, $Z=0.02$ 
star which lies on the red side of the Li dip ($T_{\rm eff} = 6220\,{\rm K}$ at 0.7~Gyr). 
For the rotation profile, IGW lead to the appearance of a ``slow''
front, propagating from the core to
the surface. When the outer convection zone is constantly spun down, several fronts propagate;
this propagation is very rapid at the beginning because differential rotation is large, and
slows down with time. 

For the evolution of chemicals, Talon \& Charbonnel (1998) expected IGW
to reduce rotational mixing together with differential rotation. 
This effect is confirmed in this first fully consistent study. 
The surface of the Li abundance in our complete model is in perfect agreement with
the data in the Hyades stars of similar effective temperature. 

Let us remind to the reader that, while the formalism presented in this paper is quite 
general and allows to properly include IGW in complete hydrodynamical stellar models,
several uncertainties remain. Firstly, a very delicate issue is that of IGW generation.
The GMK model most certainly underestimates the wave flux since it considers only bulk 
excitation, and not overshooting or convective penetration. However, since wave generation
has to be proportional to the convective luminosity, {\em differential} properties 
between different stellar types should vary less. We thus expect the wave flux to be 
larger than calculated, but by a similar amount in each stellar type. Complete studies for
different stellar types should permit to calibrate this amount, that could then be used 
uniformly.

The second uncertainty is the exact value for the turbulent
viscosity to use in the SLO ($\nu_{\rm waves}$, Eq.~\ref{nuwaves}). Here, we estimated a viscosity
based on the shear instability and assumed a small amount of enlargement and averaging in order
for the turbulent viscosity 
not to be too local. However, this model has only a rather small 
impact on numerical results since in all cases it remains very localized below the 
convection zone. 

The last point to bear in mind is that no latitudinal dependence is considered here;
only a horizontal average is used. Let us remark that, in order to be able to take into
account this dependence, observational data of latitudinal differential rotation in stars other
than the Sun must be obtained.

The results on the evolution of the rotation profile and on the lithium abundance presented
in this paper are
very encouraging and should now to be confronted to the data for stars 
of various masses, metallicities and ages. This will be done in forthcoming studies.

\begin{acknowledgements}
We would like to thank Pr. Andr\'e Maeder and Dr. Georges Meynet
for discussion on the subject of gravity wave energetics
and our referee for constructive comments.
We are grateful to the French 
``Programme National de Physique Stellaire" for financial support.
Part of the calculations were performed on computers belonging to the 
R\'eseau Qu\'eb\'ecois de Calcul de Haute Performance (RQCHP).
C.C. was supported by the Swiss National Science Foundation.
\end{acknowledgements}

\end{document}